\documentclass[fleqn,usenatbib]{mnras}

\usepackage{newtxtext,newtxmath}

\usepackage[T1]{fontenc}

\DeclareRobustCommand{\VAN}[3]{#2}
\let\VANthebibliography\thebibliography
\def\thebibliography{\DeclareRobustCommand{\VAN}[3]{##3}\VANthebibliography}

\usepackage{graphicx}	
\usepackage{amsmath}	
\usepackage{gensymb}

\title[The warped $\beta$ Pic debris disc]{Formation of the warped debris disc around $\beta$ Pictoris}

\author[J. L. Smallwood]{
Jeremy L. Smallwood\thanks{E-mail: jlsmallwood@asiaa.sinica.edu.tw} 
\\
Institute of Astronomy and Astrophysics, Academia Sinica, Taipei 10617, Taiwan\\
} 

\date{Accepted XXX. Received YYY; in original form ZZZ}

\pubyear{2023}

\begin{document}
\label{firstpage}
\pagerange{\pageref{firstpage}--\pageref{lastpage}}
\maketitle

\begin{abstract}
In light of the recent confirmation of an eccentric orbit giant planet, $\beta$ Pic c, I revisit the formation and evolution of the warped debris disc in the system. $\beta$ Pic c is interior to $\beta$ Pic b, and the  debris disc is exterior to both planets.  Previous $N$-body simulations have shown that $\beta$ Pic b is responsible for exciting the inclination of the debris disc. With hydrodynamical simulations, I model a protoplanetary gas disc misaligned with the planets.  I find that the  gas disc does not exhibit significant  long lasting inclination excitation from the planets  even for the observed disc size. The warp that is excited by the planets propagates through the entire disc with a timescale much less than the gas disc lifetime. Therefore, the observed warp in the debris disc must be produced after the gas disc has dispersed. With analytical secular theory calculations, I show that two secular resonances are exterior to $\beta$ Pic b, located at $\sim 20\, \rm au$ and $\sim 25\, \rm au$. This agrees with my $N$-body simulations that show that these secular resonances shape the inner edge of the $\beta$ Pic debris disc at a radius that agrees with observations.
\end{abstract}




\section{Introduction}

Planetary systems that house debris discs are exceptional laboratories for studying the effects of planetary dynamics. Studying the composition and structure of debris discs can lead to information on exoplanets, such as the compositions, masses, and orbits \citep{Hughes2018}. The occurrence rate of debris discs around A and F stars is estimated at $\sim 25$ per cent, while the rate is $\sim 15$ per cent around G and K type stars \citep{Su2006,Hillenbrand2008,Trilling2008,Sibthorpe2018}. Studying debris disc evolution and structure will shed light on the origin of planetary systems.

One such planetary system that has been widely studied is the $\beta$ Pictoris (Pic) system. $\beta$ Pic is a young A6V-type star with an age of $\sim 20\pm 3\, \rm Myr$ \citep{Shkolnik2012,Binks2014,Bell2015,MiretRoig2020} at a distance of 19.45 ± 0.05 pc \citep{vanLeeuwen2007}. The mass of $\beta$ Pic is estimated to be $M_{\rm star} = 1.85\pm 0.03 \, \rm M_{\odot}$ \citep{Wang2016}, with a stellar diameter of $0.736\pm 0.019\, \rm mas$ \citep{Defrere2012} and a projected rotation velocity of $v\sin i = 130\, \rm km\, s^{-1}$ \citep{Royer2007}. 

 $\beta$ Pic b was one of the first exoplanets to be discovered using direct imaging \cite[e.g.,][]{Lagrange2009,Lagrange2010}. Further observations of this planet include photometry spanning the near-infrared \citep{Quanz2010,Bonnefoy2011,Currie2011,Bonfils2013,Males2014,Morzinski2015}, low-resolution spectroscopy \citep{Chilcote2015,Chilcote2017}, and medium-resolution spectroscopy \citep{Snellen2014,GRAVITY2020}. The dynamical mass measurements of $\beta$ Pic b were derived from the observations from {\it Hipparcos} \citep{vanLeeuwen2007} and {\it Gaia} \citep{Gaia2016,Lindegren2018}. \cite{Lagrange2020} refined the mass measurements using SPHERE, GRAVITY, and RV data. Recently, the works of \cite{Nowak2020} and \cite{Brandt2021} estimate a planet mass of $~\sim 9-10\, \rm M_{Jup}$ with a semi-major axis of $\sim 10\, \rm au$. The recent confirmation of a second planet in the $\beta$ Pic system,  $\beta$ Pic c \citep{Lagrange2019,Nowak2020}, has now fueled further investigation of the planetary dynamics occurring in this system.  $\beta$ Pic c has a dynamical mass of $\sim 8\, \rm M_{Jup}$ with a semi-major axis of $\sim 3\, \rm au$ \citep{Nowak2020,Brandt2021}. Table~\ref{table::planets} shows the orbital elements for $\beta$ Pic b and $\beta$ Pic c from the  work of \cite{Brandt2021}. Both planets are on eccentric orbits with eccentricities of $\sim 0.3$ and $\sim 0.121$ for $\beta$ Pic b and $\beta$ Pic c, respectively. Moreover, the planets are coplanar to one another and viewed nearly  in an edge-on geometry.


 \begin{table*}\centering
 \caption{A summary of the planetary parameters of the $\beta$ Pic system derived by \citet{Brandt2021}.}
 \begin{tabular}{lcccc}
  \hline
  Planet Parameters & Symbol  &  $\beta$ Pictoris b &  $\beta$ Pictoris c & Unit\\
    \hline
    \hline
     \vspace{0.06cm}
    Semi-major axis & $a_{\rm p}$   & $10.31\pm 0.11$ & $2.75^{+ 0.040}_{-0.035}$ & $\rm au$ \\
    \vspace{0.06cm}
    Eccentricity axis & $e_{\rm p}$ & $0.121\pm 0.008$ & $0.30^{+ 0.20}_{-0.16}$ & -- \\
     \vspace{0.06cm}
    Inclination &  $i_{\rm p}$    & $88.94\pm 0.02$ & $89.1\pm 0.68$ & $\rm deg$\\
     \vspace{0.06cm}
    PA of ascending node & $\Omega$   & $211.95\pm 0.030$ & $211\pm 0.25$ & $\rm deg$\\
     \vspace{0.06cm}
    Argument of periastron & $\omega$    & $19.3^{+3.1}_{-3.2}$ & $114.3^{+19}_{-3.0}$ & $\rm deg$\\
     \vspace{0.06cm}
    Planet mass & $M_{\rm p}$   & $9.8^{+2.7}_{-2.6}$ & $8.3^{+1.1}_{-1.0}$ & $M_{\rm Jup}$\\
     \vspace{0.06cm}
    Period & $P_{\rm planet}$  & $24.30\pm 0.34$ & $3.348^{+0.066}_{-0.051}$ & $\rm years$\\
    \hline
 \end{tabular}
 \label{table::planets}
\end{table*}

 A bright edge-on circumstellar debris disc around $\beta$ Pic was first imaged in the mid-1980's \citep{Smith1984}. On top of this, spectroscopic observations reveal a high rate of transits of small evaporating exocomets \citep{Kiefer2014}. The presence of $\beta$ Pic b is thought to explain the infalling exocomets, given that the planet has a nonzero eccentricity \cite[e.g.,][]{Beust1996,Beust2000}. Adaptive optics coronographic images of the $\beta$ Pic disc detected the disc in the near-infrared (NIR) through scattered light down to $\sim 25\, \rm au$ from the star \citep{Mouillet1997a}. The Hubble Space Telescope (HST) also resolved the disc down to $\sim 25\, \rm au$ \citep{Burrows1995,Lecavelier1997}.  The observed outer radius of the debris disc may extend out to $\sim 1000$ of au \citep{Jason2021}. The most intriguing structural anomaly of the debris disc is a major asymmetric feature found at around $80\,  \rm au$.  The disc displays a  warp at this location where there is a $~4-5\degree$ offset between the warped inner disc and the outer main disc \citep{Kalas1995,Wahhaj2003,Weinberger2003,Golimowski2006,Nielsen2014}.
 \cite{Kraus2020} investigated the spin-orbit alignment of the planetary orbit of $\beta$ Pic b. To accomplish this, they measured the mutual inclination angle $\phi$ between the angular momentum vectors of the stellar photosphere and the planetary orbit. They found that the planetary orbit for $\beta$ Pic b is inclined by $\sim 3\degree$. Moreover, recent observations estimated that the orbit of $\beta$ Pic c is nearly coplanar to  $\beta$ Pic b \citep{Nowak2020,Brandt2021}.   Misalignment of a planet to the spin-axis of the central star can arise from  dynamical  mechanisms, including planet-planet scattering \citep{Chatterjee2008}.
 Alternatively, the protoplanetary disc around a single star may be misaligned during the planet formation stage \cite[e.g.,][]{Bate2010,Lai2011,Rogers2013,Fielding2015}.

 An inclined planet was proposed to trigger the warp structure in the debris disc \citep{Mouillet1997b}. \cite{Dawson2011} investigated the warp in the $\beta$ Pic debris disc utilizing $N$--body simulations and secular theory. At this time, only $\beta$ Pic b was known. They found that the low-inclination planet, $\beta$ Pic b, causes a forced inclination excitation within the debris disc at around $80\, \rm au$. They concluded that the inclination of the observed inner disc is solely due to $\beta$ Pic b because if $\beta$ Pic b  were aligned with the flat outer disc (which observations suggested), it would prevent another planet from creating a warp.  More recently, \cite{Dong2020} ran numerical simulations comparing the $\beta$ pic debris disc structure under the influence of one and two planets. They found that the inclusion of $\beta$ Pic C does not significantly affect the warped debris disc structure.

In this work, I  further investigate how the dynamics of the newly confirmed inner planet and the outer planet in the $\beta$ Pic system affect the debris disc. I consider the dynamics of the protoplanetary gas disc that the planets and the debris disc would have originally formed in and the dynamics of the observed debris disc.  The main point I want to address is whether the warp in the debris disc could have formed during the protoplanetary disc phase.  In Section~\ref{sec::methods}, I discuss the setup for my hydrodynamical and $N$--body simulations. In Section~\ref{sec::hydro_results}, I show the results of my  hydrodynamical simulations. In Section~\ref{sec::sec_theory}, I apply a secular resonance model to estimate the location of secular resonances,  and compare that to the $N$--body simulations.  Finally, I draw my conclusions in Section~\ref{sec::conclusions}.

\section{Numerical Methods}
\label{sec::methods}
To simulate the $\beta$ Pic system, I use both hydrodynamical and $N$--body simulations. The former is used to model the gas disc phase, and the latter is used to model the debris disc phase.

\subsection{Hydrodynamical setup}
I use the 3-dimensional smoothed particle hydrodynamics (SPH) code {\sc phantom} \citep{Price2018} to model a gas-only protoplanetary disc around $\beta$ Pic.
I consider a gas disc in the bending-wave regime, where the disc aspect ratio $H/r$ is larger than the $\alpha$ viscosity coefficient. 
The gas disc initially consists of $10^6$ equal-mass Lagrangian particles that are  distributed between the inner disc radius, $r_{\rm in} = 20\, \rm au$ and the outer disc radius, $r_{\rm out} = 200\, \rm au$.  
The inner disc radius is selected based on the observed data, in that there is little material detected at $< 25\, \rm au$ \citep{Burrows1995,Lecavelier1997}. In the simulation, material is able to flow inwards.  In order to save computational time and increase resolution, I  initially truncate the outer edge of the gas disc.  I note that the truncated outer radius maintains the angular momentum equilibrium, meaning that the outer disc still bears most of the angular momentum.   I also simulate two larger radial discs, $r_{\rm out} = 400\, \rm au$ and $r_{\rm out} = 1000\, \rm au$. To ensure that these extended disc sizes have the same resolution as the narrow disc, I increase the number of particles to $1.6\times 10^6$ for $r_{\rm out} = 400\, \rm au$  and $2.8\times 10^6$ for $r_{\rm out} = 1000\, \rm au$. Note that the observed disc size around $\beta$ Pic may be on the order of $1000\,\rm au$ \citep{Jason2021}.

The gas surface density profile is initially set as a power law distribution given by
 \begin{equation}
     \Sigma(R) = \Sigma_0 \bigg( \frac{r}{r_{\rm in}} \bigg)^{-p},
     \label{eq::sigma}
 \end{equation}
where $\Sigma_0$ is the density normalization and $p$ is the power law index. Note that the density normalization is set from the total disc mass, $M_{\rm disc}$. I select a value of $p = 3/2$. The disc mass is set at $M_{\rm disc} = 0.001\, \rm M_{\odot}$. Observational study of the CO gas in the primordial debris disc derives an upper mass of $0.01\, \rm M_{\oplus}$ \citep{Hales2019}, however, I expect the mass of the primordially gaseous circumstellar disc to be higher.  The equation of state is  locally isothermal  and scaled to the thickness of the disc with sound speed given by
 \begin{equation}
    H = \frac{c_{\rm s}}{\Omega} \propto r^{3/2-q}, 
 \end{equation}
where $\Omega = \sqrt{GM_{\rm star}/r^3}$ and $c_{\rm s}$ is the sound speed. I take $q = 0.75$ and the disc aspect ratio is set to $H/r = 0.05$ at $r = r_{\rm in}$.  The physical disc viscosity is modelled using the artificial viscosity $\alpha_{\rm AV}$, implemented in {\sc phantom} \citep{Lodato2010}, given by
 \begin{equation}
\alpha \approx \frac{\alpha_{\rm AV}}{10}\frac{\langle h \rangle}{H},
\end{equation}
where $\langle h \rangle$ is the mean smoothing length on particles in a cylindrical ring at a given radius \citep{Artymowicz1994,Murray1996,Lodato2007}.  I assume a value of disc viscosity $\alpha =0.01$, which gives $\alpha_{\rm AV} = 0.2$.   With this value of $\alpha$, the disc is resolved with a shell-averaged smoothing length per scale height of $\langle h \rangle/H \approx 0.41$.   Using the above values for $p$ and $q$ ensures that the disc is uniformly resolved, meaning that $\langle h \rangle/H$ and consequently $\alpha$ are constant over the radial extent of the disc \citep[e.g.,][]{Lodato2007}.

The  central star is modelled as a sink particle with an accretion radius of $r_{\rm acc} = 1\, \rm au$. This accretion radius is considered a hard boundary, where particles that penetrate this radius are $100$ per cent accreted, and the mass and angular momentum of the particles are added to the star. Furthermore, I include the two eccentric $\beta$ Pic planets, $\beta$ Pic b and $\beta$ Pic c. The planets have an initial semi-major axis, eccentricity, and mass given in Table~\ref{table::planets} using the observations of \cite{Brandt2021}. The planets both begin at apastron.   The semi-major axes of the planets are interior to  inner disc edge, and thus are not embedded within the disc.  However, the inner edge of the disc will subsequently viciously extend inward during the simulation and will begin interacting will the planets (see Appendix~\ref{appendix_a} for more details).   Observational estimates of the present-day tilts of the planets are $\sim 3\degree$ \cite[e.g.,][]{Kraus2020}.  Planets on inclined orbits will damp back to coplanar due to viscous interactions with the disc \cite[e.g.,][]{Tanaka2002}.  I simulate two scenarios where the initial mutual inclination between the planets' and disc's angular momentum vectors are $\gamma_{\rm p} =10^\circ$ and $\gamma_{\rm p} =20^\circ$, which are larger than the present-day misalignment.
 The planetary accretion radius is equal to the central star to speed up computational time.
  When $\langle h \rangle/H > 0.5$, the simulation in considered unresolved. Therefore, I stop the simulation once  $\langle h \rangle/H \approx 0.5$  at the inner disc edge, which equates to a run time of $\sim 6000\, \rm P_{b}$, where $\rm P_{b}$ is the orbital period of planet b.
 Note that $1\, \rm P_{b} \approx 24\, \rm yr$, giving a total simulation time of $\sim 144,000\, \rm yr$, which is much shorter than the average  gas disc lifetime. 

 To analyse the data from the SPH simulations, I divide the protoplanetary gas disc into $200$ bins in spherical radius, $r$, which ranges from $10\,\rm au$ to $250\, \rm au$ for $r_{\rm out} = 200$, $10\,\rm au$ to $450\, \rm au$ for $r_{\rm out} = 400$, and $10\,\rm au$ to $1100\, \rm au$ for $r_{\rm out} = 1000$. Within each bin, I calculate the particles' mean properties, such as the surface density, inclination, eccentricity, and longitude of the ascending node, in the centre-of-mass frame. The inclination and longitude of the ascending node are measured relative to the system's total angular momentum. The inclination is denoted as $i$, and the average initial misalignment of an object's angular momentum vector to the total angular momentum vector of the system is denoted as $i_0$. To compare the results of the simulations, I measure the quantity $i/i_0$. An object is considered coplanar when its angular momentum vector is aligned with the total angular momentum vector, $i/i_0 = 0$.



\subsection{$N$--body simulation setup}
To model the evolution of a debris disc, I use the hybrid symplectic integrator in the orbital dynamics package, {\sc mercury}, which uses $N$--body integrations to calculate the orbital dynamics of objects in the gravitational potential of a star \citep{Chambers1999}. I simulate the motion of the eccentric giant planets, $\beta$ Pic b and $\beta$ Pic c, and a distribution of test particles orbiting $\beta$ Pic. The test particles only interact gravitationally with the planets and the central star. I can model the debris disc as a population of test particles and neglect the particle-particle interactions because some of the largest asteroids have collisional timescales that are of the order of the age of the Solar system \citep{Dohnanyi1969}. I calculate the system evolution for a  duration of $20\, \rm Myr$ or $\sim 870,000\, \rm P_{b}$, which is equivalent to the age of the system \citep{MiretRoig2020}.

The orbits of $\beta$ Pic b and $\beta$ Pic c are taken from Table~\ref{table::planets}, using the parameters from \cite{Brandt2021}. Therefore, the planets are inclined by roughly $3\degree$ with respect to the debris disc. The debris belt contains $10,000$ test particles with a semi-major axis, $a$, randomly  distributed between $20\, \rm au$ to $200\,\rm au$. I select a disc aspect ratio of $\sim 1^\circ$ so that the disc opening angle is initially less than the initial planetary tilts. Each particle initially begins with an eccentricity randomly allocated between $0$ to $0.025$. The remaining orbital elements, the longitude of the ascending node ($\Omega$), the argument of perihelion ($\omega$), and the mean anomaly ($M_{\rm a}$) are all randomly allocated in the range $0 - 360\degree$.  The orbital elements from the $N$--body simulations are measured in the same fashion as the SPH simulations for consistency.



\begin{figure*}
\includegraphics[width=\columnwidth]{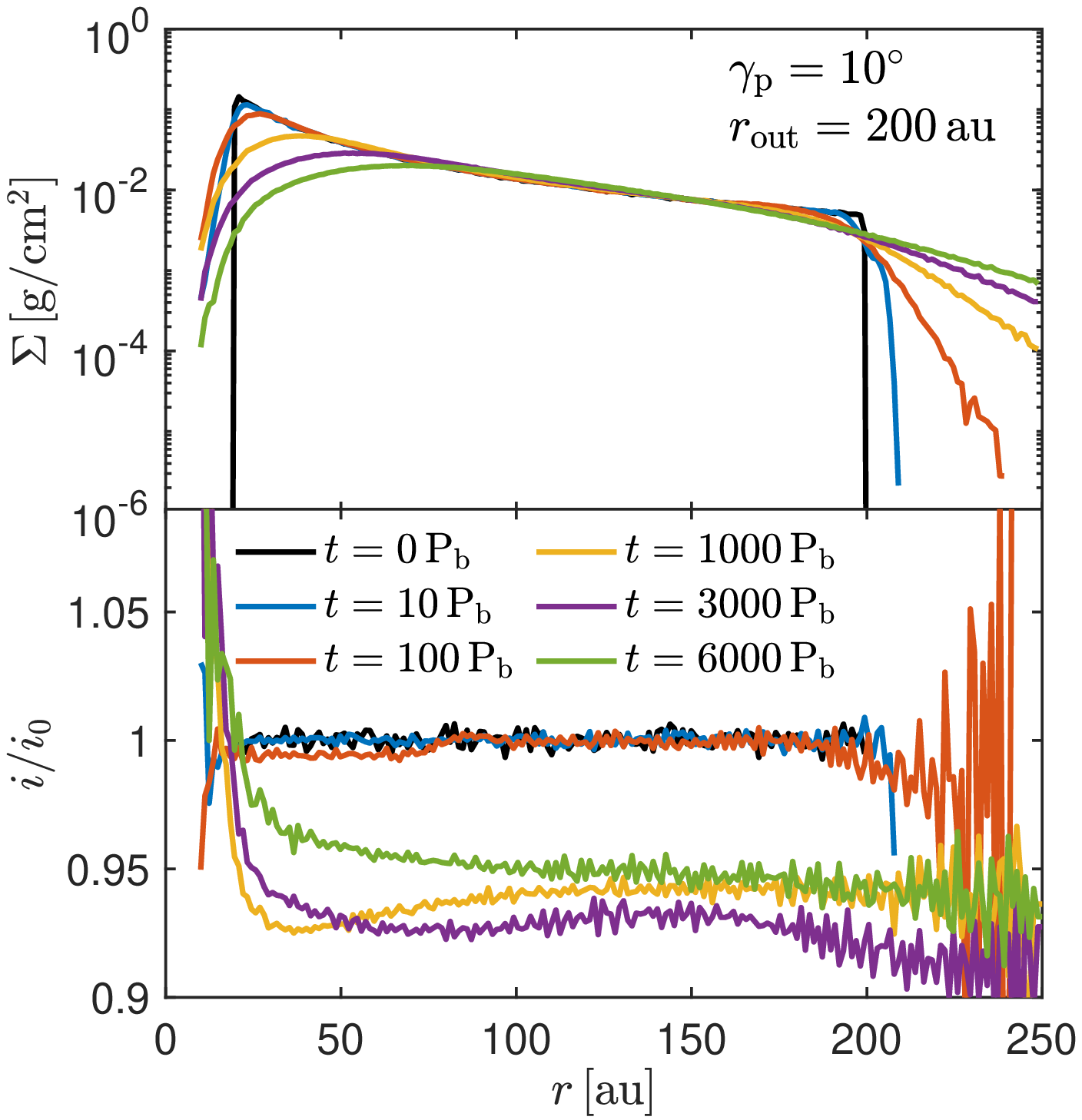}
\includegraphics[width=\columnwidth]{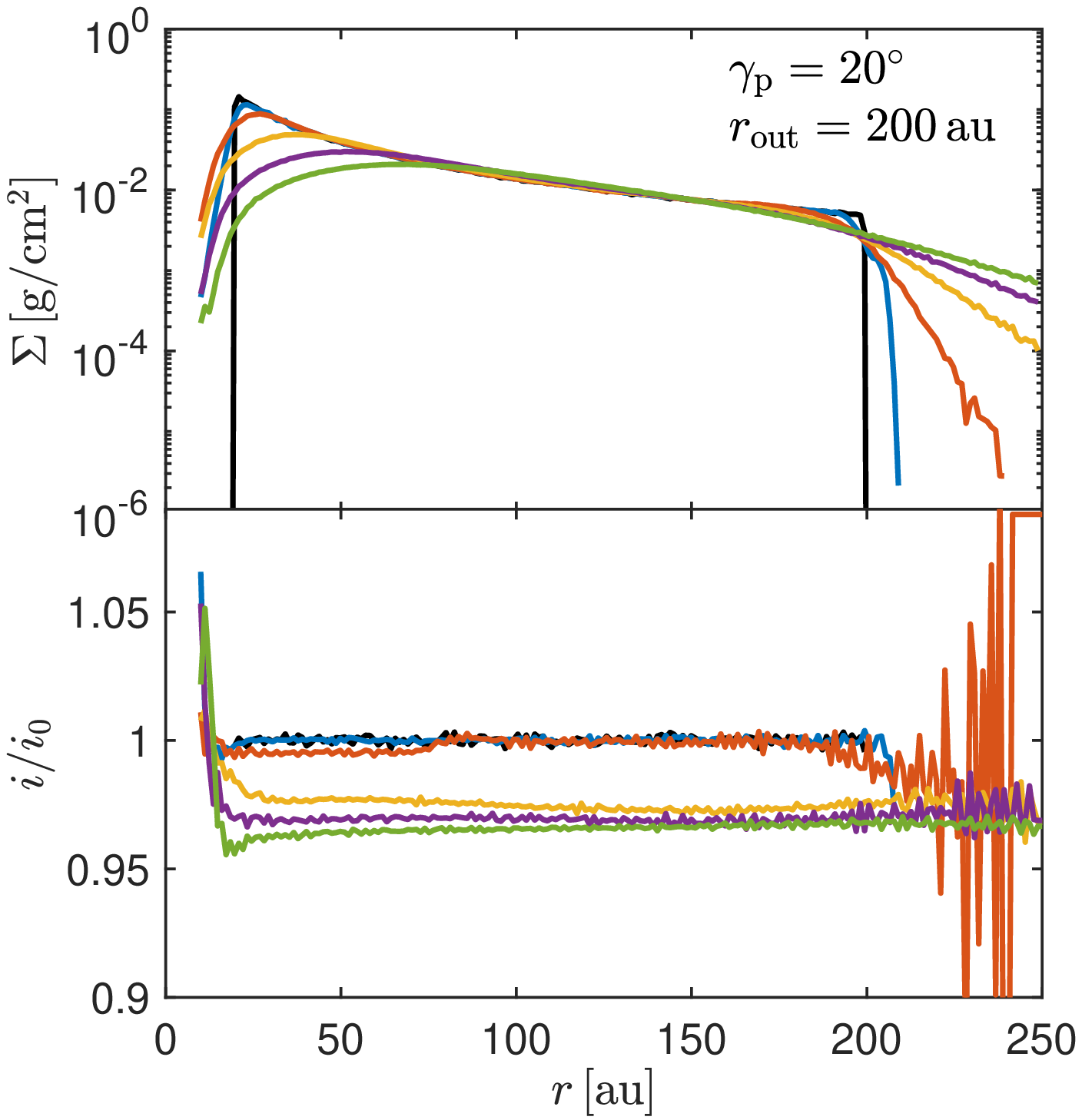}
\includegraphics[width=\columnwidth]{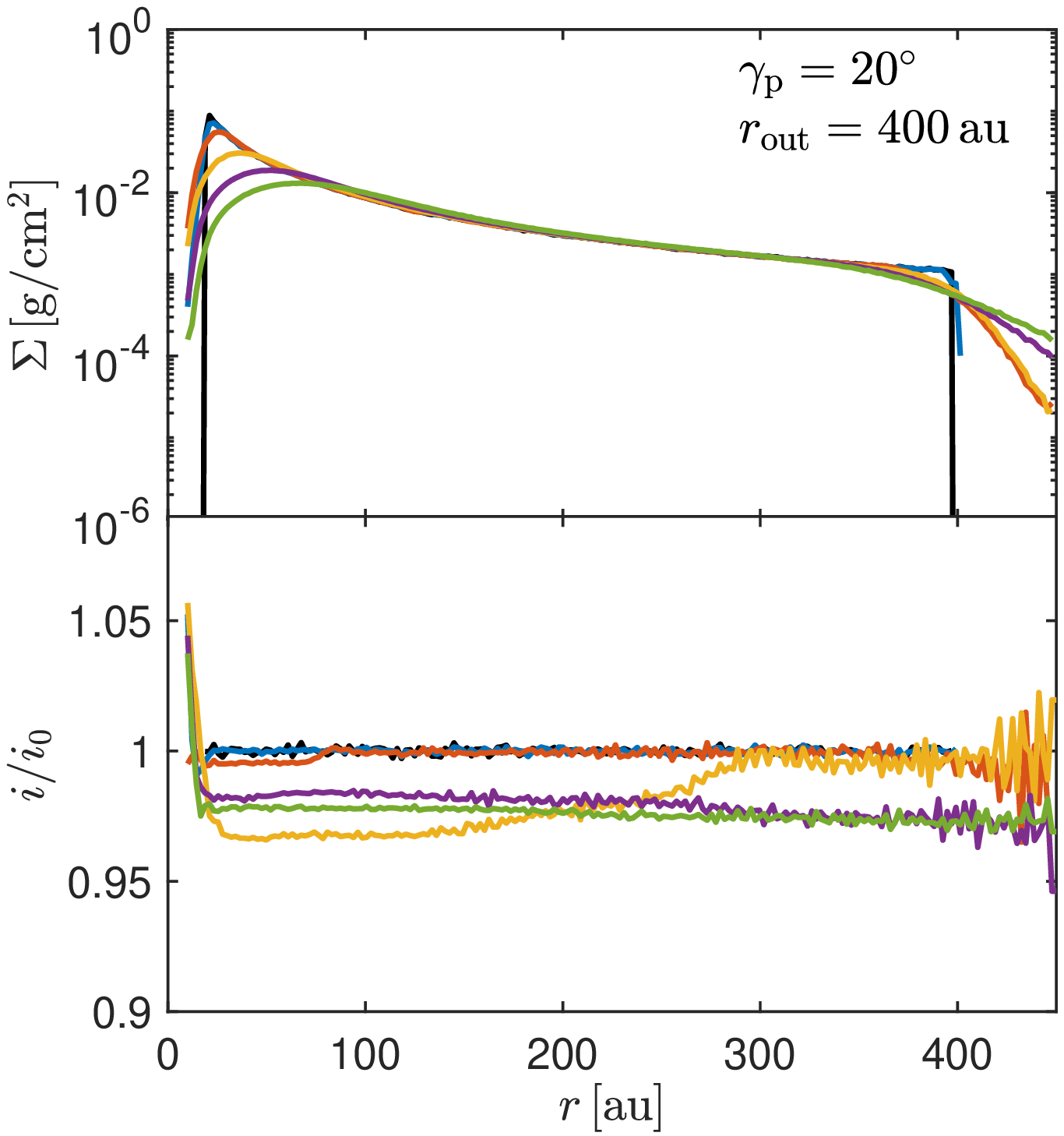}
\includegraphics[width=\columnwidth]{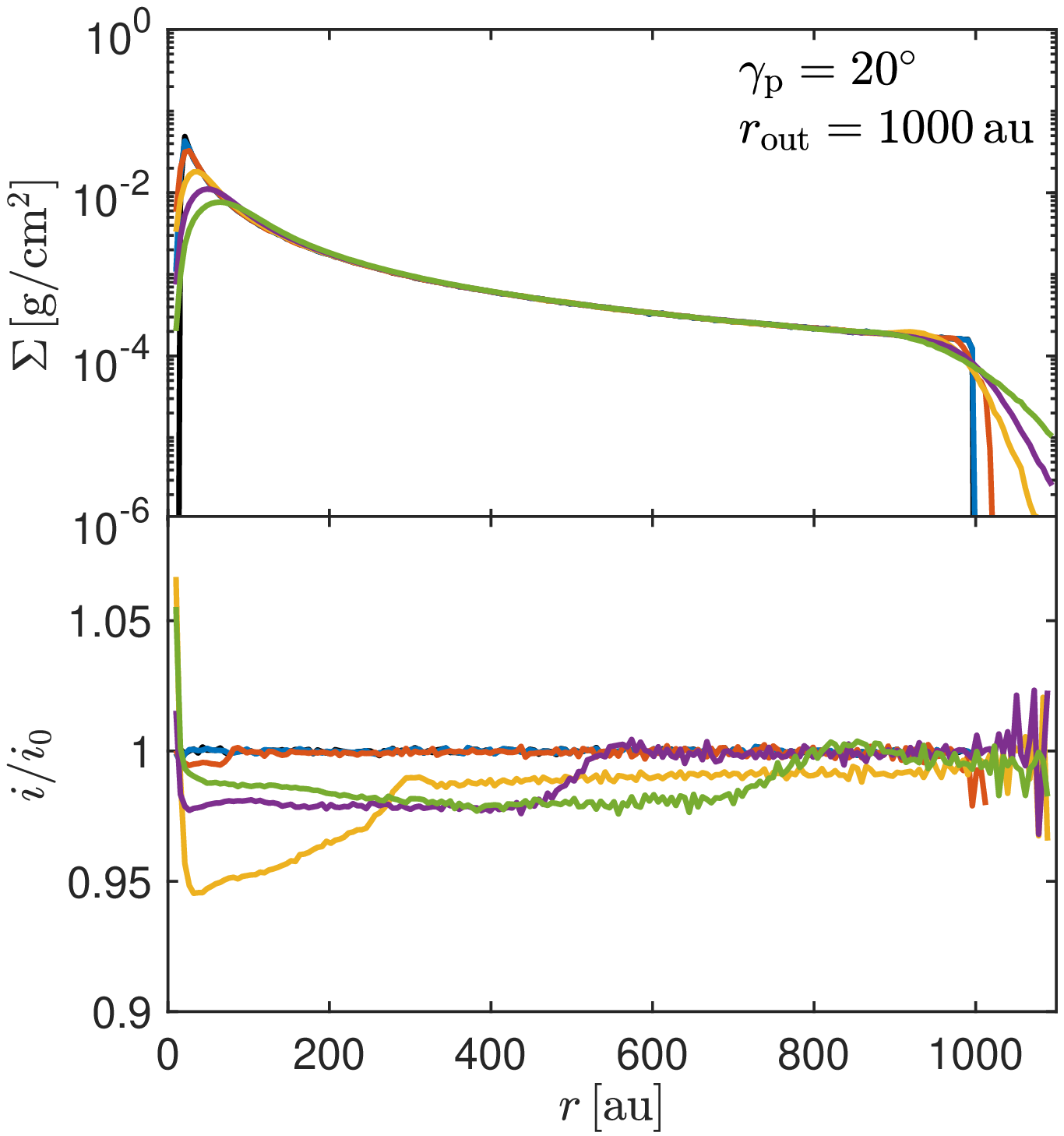}
\caption{Evolution of the gas disc from the hydrodynamic simulations.    Top-left panel: initial planetary tilt $\gamma_{\rm p} = 10^\circ$ with $r_{\rm out} = 200\, \rm au$. Top-right panel: initial planetary tilt $\gamma_{\rm p} = 20^\circ$ with $r_{\rm out} = 200\, \rm au$. Bottom-left panel: initial planetary tilt $\gamma_{\rm p} = 20^\circ$ with $r_{\rm out} = 400\, \rm au$. Bottom-right panel: initial planetary tilt $\gamma_{\rm p} = 20^\circ$ with $r_{\rm out} = 1000\, \rm au$. The surface density ($\Sigma$, upper sub-panel) and tilt ($i/i_0$, bottom sub-panel) as a function of radius, $r$. The  black lines denote the disc parameters at $t=0\, \rm P_{\rm b}$, where $P_{\rm b}$ is the orbital period of the outer-most planet ($\beta$ Pic b). I show six different times given by the legend in the top-left panel. The gas disc does not show a long lasting warp in the disc tilt.}
\label{fig::disc_params}
\end{figure*}

\section{Hydrodynamical simulations}
\label{sec::hydro_results}
 I explore the evolution of a protoplanetary disc around $\beta$ Pic, along with the two observed eccentric orbit giant planets. I consider two initial planetary tilts, $\gamma_{\rm p} = 10^\circ$ and $\gamma_{\rm p} = 20^\circ$, with various disc sizes. Below, I describe the disc and warp structure from the hydrodynamical simulations.

\begin{figure*}
\includegraphics[width=\columnwidth]{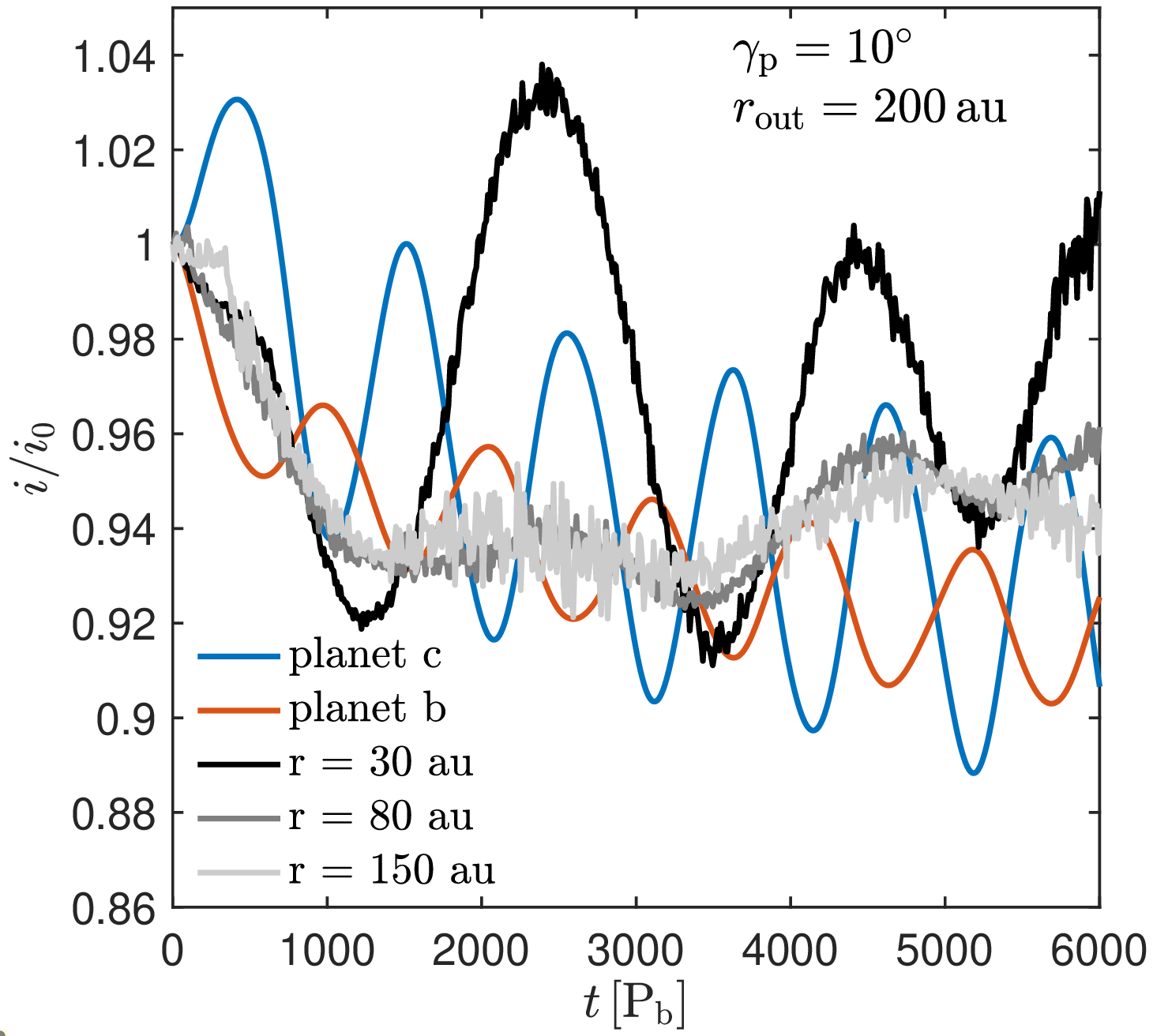}
\includegraphics[width=\columnwidth]{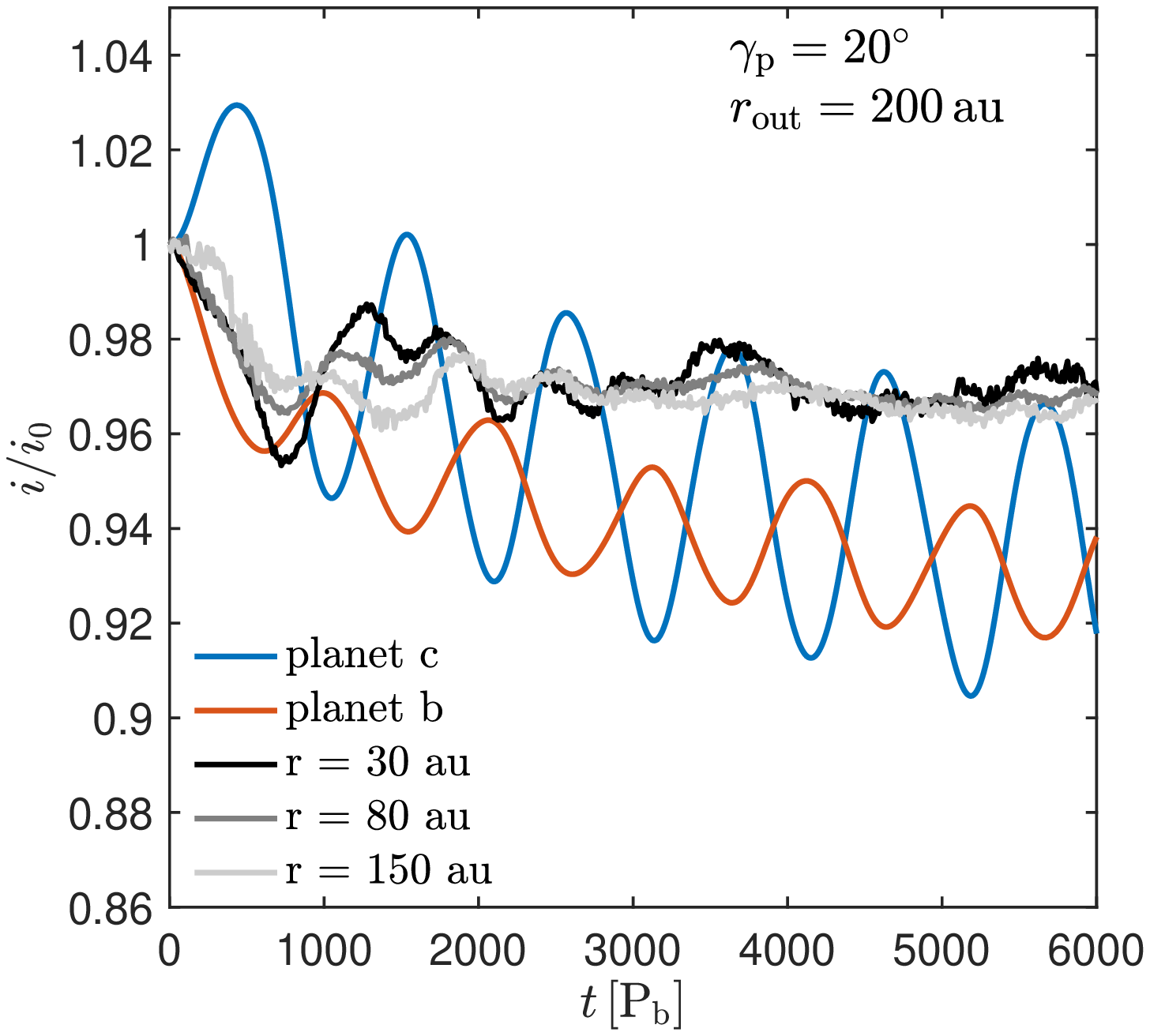}
\includegraphics[width=\columnwidth]{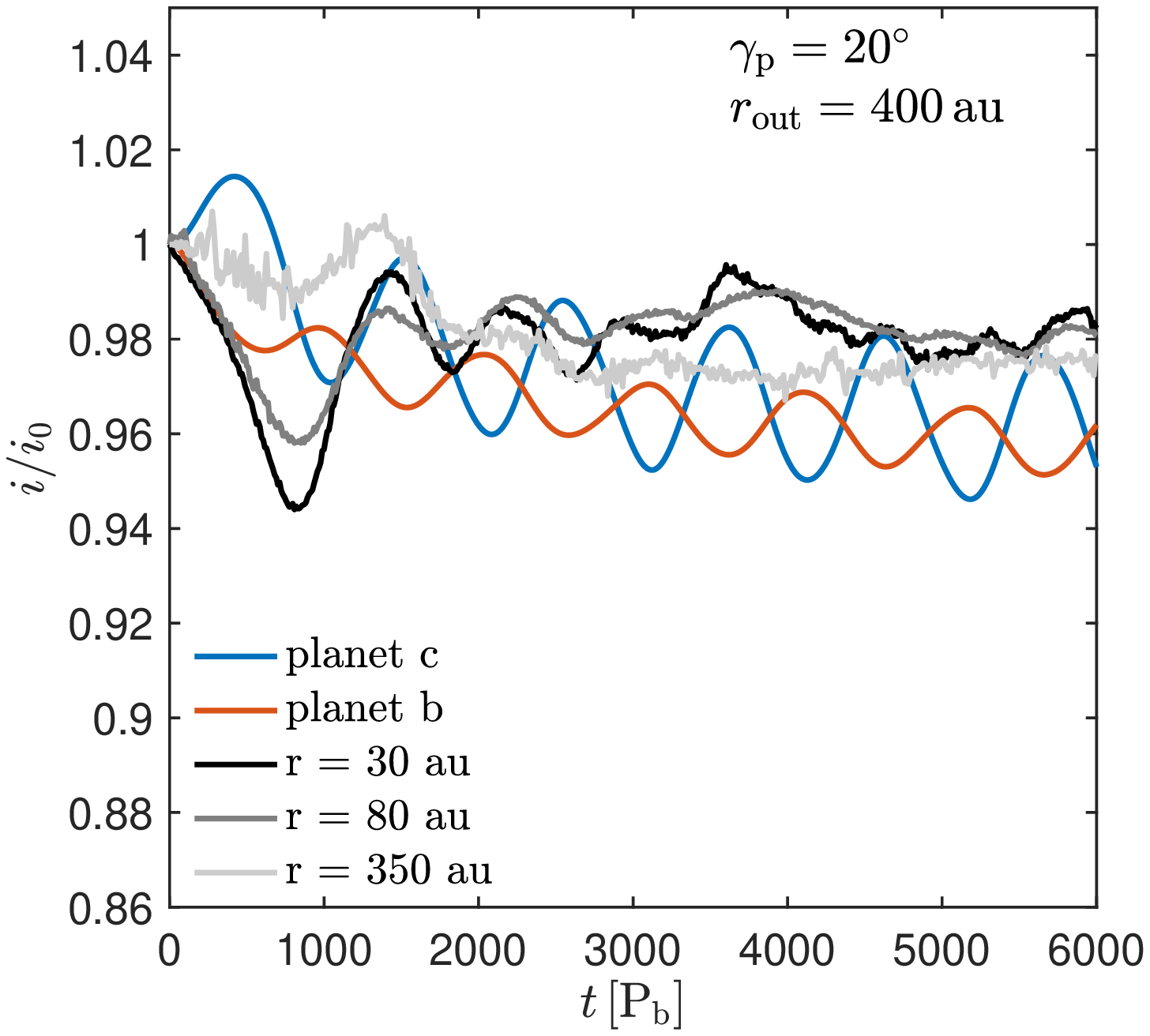}
\includegraphics[width=\columnwidth]{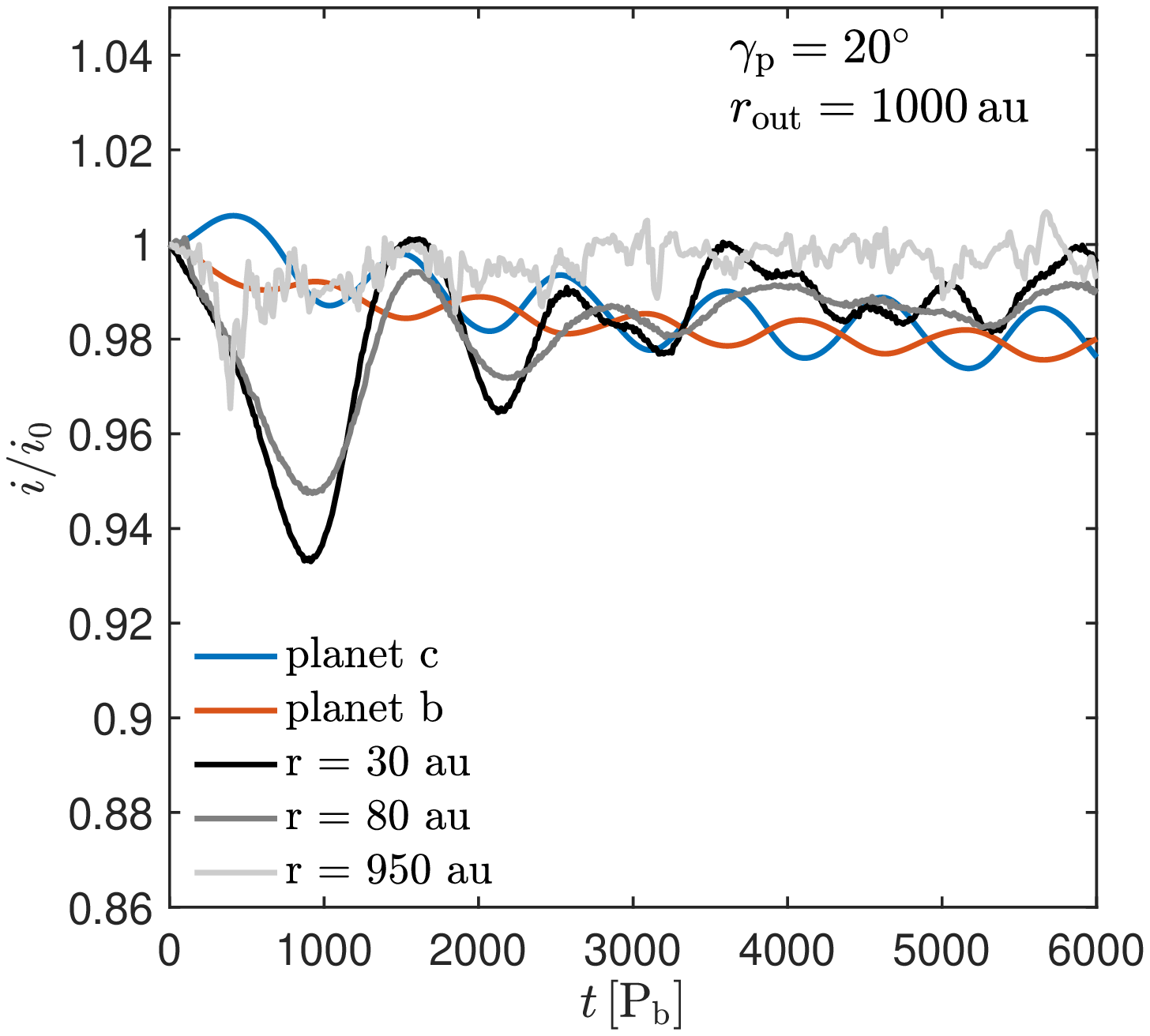}
\caption{ The tilt, $i/i_0$, evolution for the $\beta$ Pic planets and disc from each hydrodynamical simulation.   Top-left panel: initial planetary tilt $\gamma_{\rm p} = 10^\circ$ with $r_{\rm out} = 200\, \rm au$. Top-right panel: initial planetary tilt $\gamma_{\rm p} = 20^\circ$ with $r_{\rm out} = 200\, \rm au$. Bottom-left panel: initial planetary tilt $\gamma_{\rm p} = 20^\circ$ with $r_{\rm out} = 400\, \rm au$. Bottom-right panel: initial planetary tilt $\gamma_{\rm p} = 20^\circ$ with $r_{\rm out} = 1000\, \rm au$.
The tilt of $\beta$ Pic c (blue) and $\beta$ Pic b (red) as a function of time $\rm P_{\rm b}$, where $\rm P_{\rm b}$ is the orbital period of the outer-most planet ($\beta$ Pic b).   I show the disc tilt at three different radii in order of decreasing color shade given by the legend.  I probe the inner region, $80\, \rm au$, and outer region of the disc for each simulation. The observed warp is located at $80\, \rm au$.}
\label{fig::time}
\end{figure*}

\subsection{Disc structure}
 I first analyse the simulations with initial planetary tilts  $\gamma_{\rm p} = 10^\circ$ and $20^\circ$ with an outer disc radius $r_{\rm out} = 200\, \rm au$. The top left and right panels in  Fig.~\ref{fig::disc_params} show the evolution of the disc surface density $\Sigma$ and tilt $i/i_0$ for $\gamma_{\rm p} = 10^\circ$ and $\gamma_{\rm p} = 20^\circ$, respectively. I show the disc structure at times $t=0\,, 10\,, 100\,, 1000\,, 3000\,, 6000\, \rm P_{\rm b}$, where $P_{\rm b}$ is the orbital period of $\beta$ Pic b. Over time, the surface density profile slowly evolves. The inner part of the disc accretes onto the central star, and the outer portion of the disc viscously spreads outwards.  In each scenario, a warp is generated in the inner disc region seen at $t = 100\, \rm P_{\rm b}$, then the warp propagates across the entire disc. After the warp dissipates, the disc maintains a coherent flat structure throughout the simulation, seen at $t = 6000\, \rm P_{\rm b}$.  The tilt of the disc gradually aligns with the system's total angular momentum. 
 There is no evidence of a  long-lasting warp produced by the inclined planets during the gas disc phase.

Next, I examine the simulation with a more extended disc outer edge, $r_{\rm out} = 400\, \rm au$ and $1000\, \rm au$. Note that  $r_{\rm out} = 1000\, \rm au$ is the observed outer disc radius \citep{Jason2021}. In these simulations, the planets' have an initial tilt $\gamma_{\rm p} = 20^\circ$. The evolution of the surface density $\Sigma$ and tilt $i/i_0$ are shown in the bottom left and right panels in Fig.~\ref{fig::disc_params} for $r_{\rm out} = 400\, \rm au$ and $1000\,\rm au$, respectively. Again, I show the disc structure at times $t=0\,, 10\,, 100\,, 1000\,, 3000\,, 6000\, \rm P_{\rm b}$. For $r_{\rm out} = 400\, \rm au$, the surface density profile slowly evolves, where the inner part of the disc accretes onto the central star, and the outer portion of the disc viscously spreads outwards. Similar to the narrow disc simulation, a warp is generated within the disc, seen at $t = 1000\, \rm P_{b}$. At $t = 3000\, \rm P_{b}$, the warp has propagated across the entire disc, forcing the disc to evolve as a rigid body. 
For $r_{\rm out} = 1000\, \rm au$, the disc surface density evolves in a similar fashion, where the inner regions of the disc accrete onto the central star, while the outer regions of the disc viscously spread outwards over time. The warp that is excited in the disc remains for the duration of the simulation. However, the warp propagates further than $80\, \rm au$, which is the present location of the observed warp. I estimate the warp propagation timescale for this particular disc size in the following section.


The top left and right panels in Fig.~\ref{fig::time} show the planetary tilt as a function of time, as well as the disc tilt at three different radii for the truncated disc simulations ($r_{\rm out} = 200\, \rm au$). The evolution of the remaining orbital elements for the $\beta$ Pic planets for each simulation are given in Appendix~\ref{appendix_c}.  The tilt of the planets oscillates in time, driven by the interaction with the gas disc. For the truncated disc simulations, I probe the disc at radii  $r = 30\, \rm au$ (inner disc region), $80\, \rm au$ (observed warp region), $150\, \rm au$ (outer disc region).  The planets and gas disc begin to align to the total angular momentum of the system. 
For each simulation, the inner and outer disc regions align on the same timescale at $6000\, \rm P_{orb}$, indicating that the disc is not warped.

The bottom left and right panels in Fig.~\ref{fig::time} show the planetary and disc tilt as a function of time for the more extended disc simulations, $r_{\rm out} = 400\, \rm au$ and $1000\, \rm au$. For $r_{\rm out} = 400\, \rm au$, the disc is warped for a longer period of time, but eventually, the whole disc undergoes alignment as a rigid body. For $r_{\rm out} = 1000\, \rm au$, the disc can maintain a warp within the simulation time domain, but the warp has propagated past $80\, \rm au$.  The torque applied by the planets is the same between each simulation, regardless of disc size. Since the torque and angular momentum are related by a rate of change, applying the same torque to a disc with different angular momentum will result in a different alignment timescale. Therefore, it is expected that the disc with the larger angular momentum takes longer to align, which is consistent with the results displayed in Figs.~\ref{fig::disc_params} \&~\ref{fig::time}. I only simulate one value of the disc aspect ratio $H/r$, but for larger values of $H/r$, the disc will align on a faster timescale \cite[e.g.,][]{Lubow2018}.

\begin{figure}
\includegraphics[width=1\columnwidth]{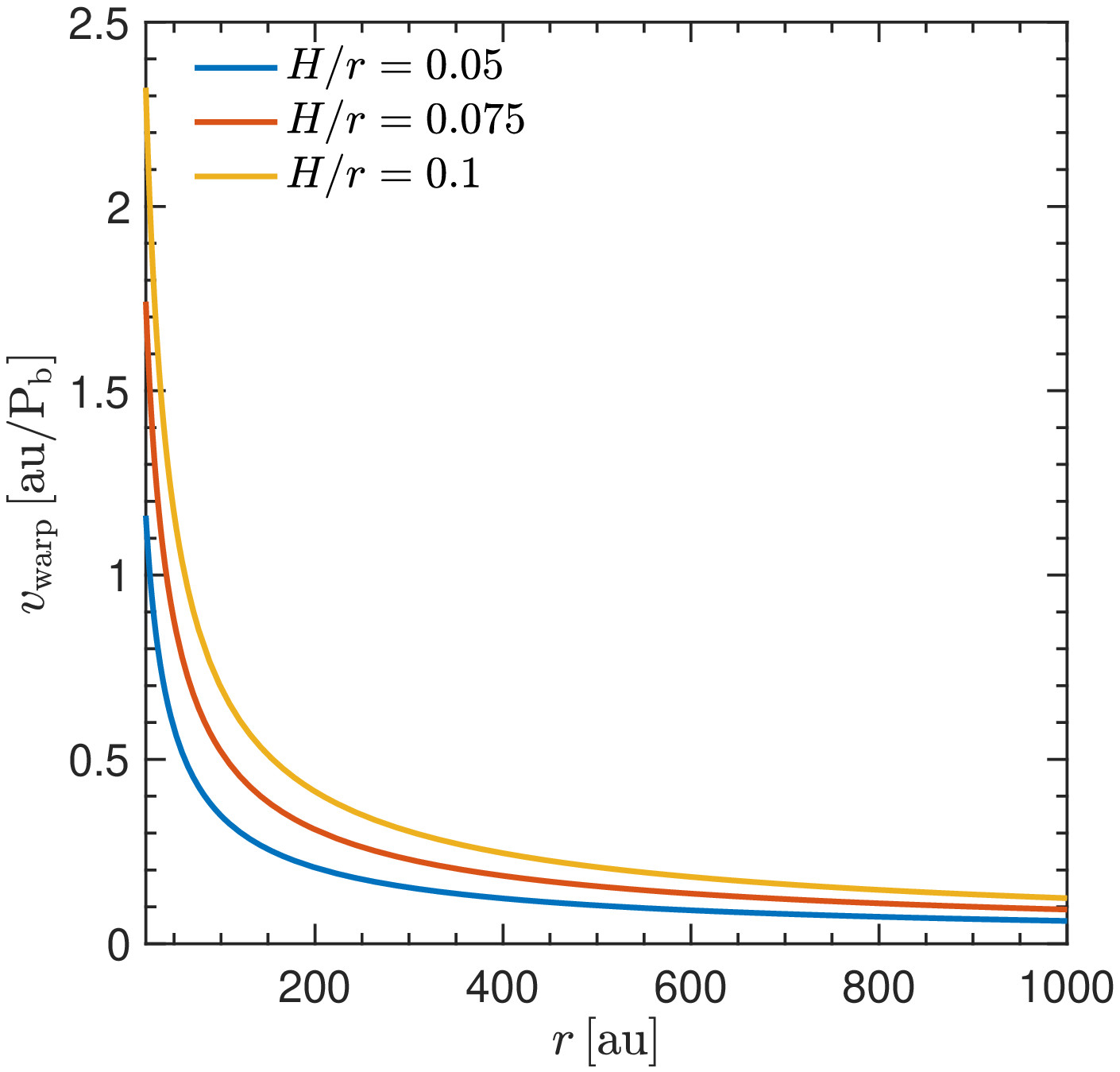}
\caption{ The warp propagation velocity, $v_{\rm warp}$, from Eq.~(\ref{eq::vwarp}) as a function of time in orbital periods of the outer-most planet, $\rm P_{b}$. The warp propagation velocity is computed using three different values of the disc aspect ratio, $H/r = 0.05$ (blue), $0.075$ (red), and $0.1$ (yellow). Note that for the hydrodynamical simulations, $H/r = 0.05$.}
\label{fig::warp}
\end{figure}

\begin{figure}
\includegraphics[width=1\columnwidth]{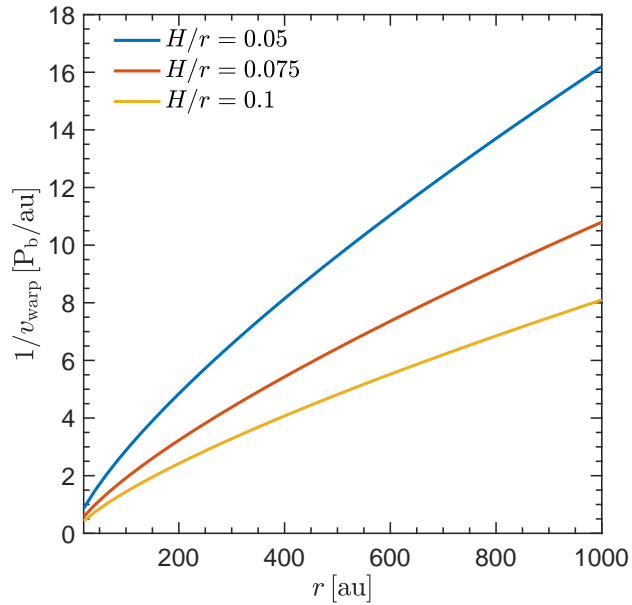}
\caption{Similar to Fig.~\ref{fig::warp}, except showing the inverse of the warp propagation velocity.}
\label{fig::warp_inv}
\end{figure}

\begin{figure}
\includegraphics[width=1\columnwidth]{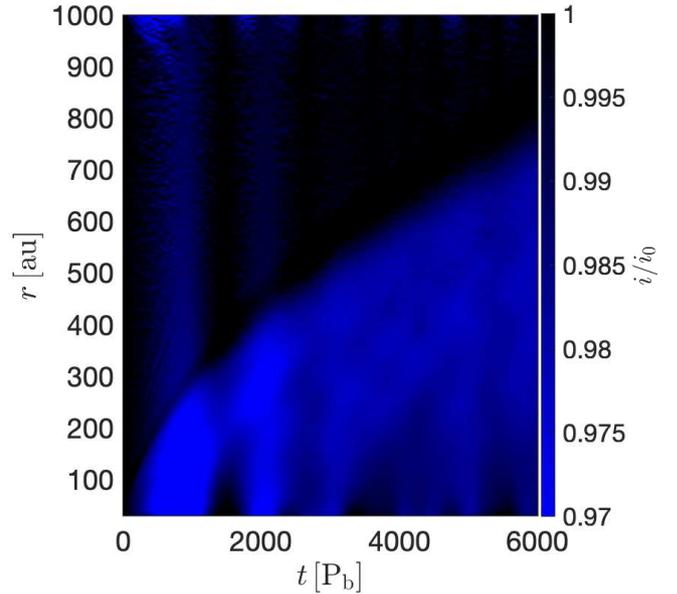}
\caption{ The disc tilt evolution for the observed disc size ($r_{\rm out} = 1000\, \rm au$). The $x$ and $y$ axes give the time in orbital period of the outer-most planet ($\beta$ Pic b), $P_{\rm b}$, and disc radius, $r$, respectively. The colour bar  denotes the disc tilt, $i/i_0$.  Over time, the disc aligns to the total angular momentum of  the system as the warp induced by the planets propagates outward in the disc.}
\label{fig::time_tilt_r1000}
\end{figure}

\subsection{Warp propagation}
 Two regimes govern the propagation of warps, the bending-wave regime ($\alpha \lesssim H/R$) and the diffusion regime ($\alpha \gtrsim H/R$) \citep{Papaloizou1983}. In the case of my hydrodynamical simulations of a protoplanetary disc around $\beta$ Pic, the misalignment between the planets and disc will induce a warp that travels from the inner disc edge to the outer disc edge. Since my simulations are in the regime where $\alpha \lesssim H/R$, the warp propagates as a bending wave with a propagation speed of half the sound speed \citep{Papaloizou1995}. For the simulations with an initial planetary tilt of $20^\circ$, I estimate the warp propagation timescale for the different disc sizes, $r_{\rm out} = 200\, \rm au$, $400\, \rm au$, and $1000\, \rm au$. 
  The sound speed, $c_{\rm s}$ is given by
\begin{equation}
    c_{\rm s} = \frac{H}{r}\bigg\rvert_{r = r_{\rm in}} \sqrt{\frac{G M_{\rm star}}{r_{\rm in}}} \bigg( \frac{r}{r_{\rm in}}\bigg)^{-q},
\end{equation}
where $H/r$ is the disc aspect ratio  evaluated at the initial inner disc edge, $G = 4 \pi^2 \, \rm au^3 yr^{-2} M_{\odot}$ is the gravitational constant, $M_{\rm star} = 1.85\, \rm M_{\odot}$ is the central star mass, $r$ is the disc radius, and $r_{\rm in} = 20\, \rm au$ is the initial inner disc edge. The warp propagation velocity is then
\begin{equation}
    v_{\rm warp} = \frac{c_{\rm s}}{2}.
    \label{eq::vwarp}
\end{equation}

Figure~\ref{fig::warp} shows the warp propagation velocity from Eq.~(\ref{eq::vwarp}) as a  function of radius for three different disc aspect ratio values, $H/r = 0.05$ (used in the hydrodynamical simulations), $0.075$, and $0.1$. The warp propagates faster near the inner edge of the disc, then slows down as the warp approaches the outer disc edge. The warp will have a higher propagation velocity for a disc with a larger disc aspect ratio. Figure~\ref{fig::warp_inv} shows the inverse of the warp propagation velocity as a  function of radius for the three different disc aspect ratio values. The time it takes for the warp to propagate to a given radius can be determined by integrating the curve.  Using $H/r = 0.05$, the time it takes the warp to propagate to $200, \rm au$, $400\, \rm au$, and $1000\, \rm au$ is $\sim 543\, \rm P_{b}$, $\sim 1852\, \rm P_{b}$, and $\sim 9245\, \rm P_{b}$, respectively.  The warp will propagate across the entire disc within the gas disc lifetime, even for the observed disc size of $1000\, \rm au$, which is consistent with the results from Figs~\ref{fig::disc_params} \&~\ref{fig::time}. A more detailed look at the tilt for this disc size ($r_{\rm out} = 1000\, \rm au$) is given in Figure~\ref{fig::time_tilt_r1000}, which shows the disc tilt evolution as a function of time on the $x$--axis and disc radius on the $y$--axis. The warp is still propagating outward at $6000\, \rm P_{b}$ but has fully propagated beyond the observed warp radii of $80\, \rm au$. This means the outermost planet likely produces the observed warp during the debris disc phase.

\subsection{Limitations}

 By definition, in SPH, all the material accreted by a sink particle is added to the planet's mass. However, in actual fact, a fraction of the accreted mass may end up in an unresolved circumplanetary disc that aids in softening the accretion of material onto the planet. Therefore, the computed planetary mass evolution will be an upper limit. Moreover, since the planets in the hydro simulations are initially interior to the inner disc edge, the region around the planets will be unresolved due to the lack of SPH particles viscously drifting inwards. The unresolved accretion flow onto the planets may impact how quickly the planetary orbits damp to coplanar.

\section{$N$--body and Secular Resonance Models}
\label{sec::sec_theory}
 In the previous section, I found that the inclined $\beta$ Pic planets do not excite a lost-lasting warp in the gaseous protoplanetary disc. Therefore, the warp observed in the debris disc should be produced after the gas disc has dispersed. In this section, I further investigate the dynamics of the warp during the debris disc phase utilizing $N$--body integrations with updated orbital parameters for $\beta$ Pic b and the newly confirmed inner planet, $\beta$ Pic c. In addition, I determine the location and strength of apsidal secular resonances in the $\beta$ Pic system. I first consider the apsidal eigenfrequency of each planet and then find the free precession rate of a test particle in the system. The location where the free precession rate is equal to an eigenfrequency is an apsidal resonance location. Finally, I calculate the forced eccentricity of a test particle. The analytical model I use is linear in eccentricity and inclination and calculates the secular perturbations to second order in eccentricity.  I then compare the secular theory results to the $N$--body simulations.

\subsection{$N$--body Simulations}
\begin{figure}
 \includegraphics[width=\columnwidth]{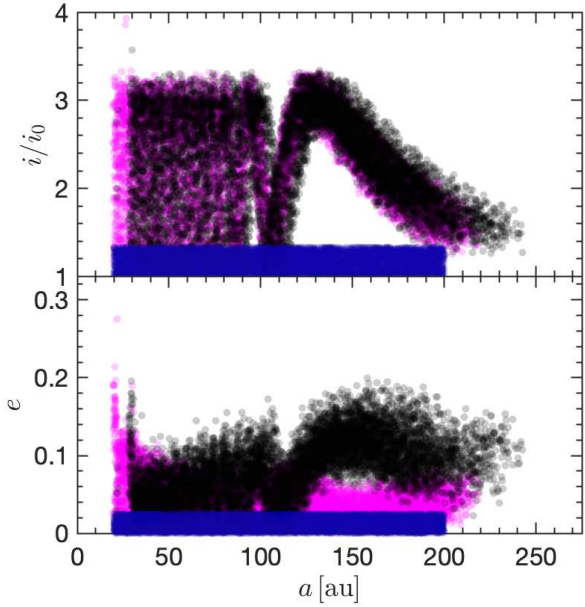}
 \caption{ Top panel: The inclination, $i/i_0$, as a function of semi-major axis, $a$, for the $N$-body test particles. Bottom panel: The eccentricity, $e$, as a function of $a$. The blue dots show the initial distribution. The magenta and black dots show the distribution at $t = 20\, \rm Myr$ with one planet ($\beta$ Pic b), and two planets ($\beta$ Pic b and $\beta$ Pic c), respectively.   }
 \label{fig::nbody}
 \end{figure}

 Figure~\ref{fig::nbody} shows the inclination (upper panel) and eccentricity (lower panel) distributions of test particles as a function of semi-major axis. I compare the results of the simulation between one planet ($\beta$ Pic b, magenta dots) and two planets ($\beta$ Pic b + $\beta$ Pic c, black dots). The  warp is present when using the updated orbital parameters of $\beta$ Pic b from \cite{Brandt2021}, and is consistent with the orbital parameters in the simulations conducted by \cite{Dawson2011}. I obtain a similar warp structure when I include the inner planet, $\beta$ Pic c. The width of the extended outer tail of the distribution is similar in both cases. Recently, \cite{Dong2020} ran numerical simulations comparing the $\beta$ Pic debris disc structure under the influence of one and two planets. They found that the inclusion of the inner planet does not significantly affect the warped debris disc structure. However, the previous works did not look at the eccentricity growth in the debris disc. When the two-planet system is modelled, there is additional eccentricity growth in the outer regions of the disc that is not present when only the outer planet is simulated. Moreover, the inner edge of the disc is truncated when two planets are present versus only the outer-most planet is modelled. In the following subsections, I propose that secular resonances are the reason for the truncation of the inner edge.

\subsection{Apsidal Eigenfrequency}
Here, I calculate the apsidal eigenfrequencies of a planetary system with a total of $N$ planets orbiting a star with mass $m_*$. I write my equations generally, but for the model of $\beta$ Pic I take $N=2$ and $m_* = 1.85 \, \rm M_{\odot}$ \citep{Wang2016}.   Each planet has  semi-major axis $a_j$, eccentricity $e_j$, mass $m_j$, longitude of the perihelion $\bar{\omega}_j$ (which is defined as the addition of the argument of pericenter ($\omega_j$) and the longitude of the ascending node ($\Omega_j$)),  and orbital frequency $n_j = \sqrt{Gm_* / a_j^3}$, where $j=1,...,N$.

The apsidal eigenfrequency of each planet is found by calculating the eigenvalues of the $N\times N$ matrix $A_{jk}$ associated with the generalized form of the secular perturbation theory
\begin{equation}
A_{jk} = -\frac{1}{4} \frac{m_k}{m_* + m_j}n_j \alpha_{jk}\bar{\alpha}_{jk}b^{(2)}_{3/2}(\alpha_{jk}) 
\label{A1}
\end{equation}
for $j\ne k$ and otherwise
\begin{equation}
A_{jj} = \frac{n_j}{4} \sum_{k=1,k\neq j}^{N} \frac{m_k}{m_* + m_j}\alpha_{jk}\bar{\alpha} _{jk}b^{(1)}_{3/2}(\alpha_{jk})
\label{A2}
\end{equation}
\citep{MurrayBook2000,Minton2011,malhotra2012,Smallwood2018a,Smallwood2018b,Smallwood2021}, where the Laplace coefficient $b_{s}^{(j)}(\alpha)$ is given by
\begin{equation}
\frac{1}{2}b_{s}^{(j)}(\alpha) = \frac{1}{2\pi}\int_0^{2\pi}\frac{\cos (j\psi)\, d\psi}{(1-2\alpha \cos \psi + \alpha^2)^s}
\end{equation}
and the coefficients $\alpha_{jk}$ and  $\bar{\alpha}_{jk}$ are defined as
\begin{equation}
  \alpha_{jk}=\begin{cases}
    a_k/a_j, & \text{if $a_j > a_k$},\\
    a_j/a_k, & \text{if $a_j < a_k$},
  \end{cases}
\end{equation}
and
\begin{equation}
  \bar{\alpha}_{jk}=\begin{cases}
    1, & \text{if $a_j > a_k$},\\
    a_j/a_k, & \text{if $a_j < a_k$}.
  \end{cases}
\end{equation}

\begin{table}
	\centering
	\caption{The values derived from the first-order linear secular theory model. Column $2$ shows the apsidal secular eigenfrequency ($g_i$) for each of the planets in $\beta$ Pic. Columns 3 and 4 show the $e_{ji}$ components for $j=1-2$ of the apsidal eigenvectors for the eigenvalue solutions from the matrix $A_{jk}$ derived from the apsidal secular perturbation theory for the $\beta$ Pic planetary system. Column $5$ denotes the phase angle $\beta_i$.}
	\begin{tabular}{ccccc} 
		\hline
		Planet &  $g_i$ & \multicolumn{2}{c}{$e_{\rm ji}$}  & $\beta_i$ \\
        &  ($^{\prime \prime} \rm yr^{-1}$) & $j=1$ & $2$ & (deg.)   \\
		\hline
		$\beta$ Pic c ($i=1$) & $15.7146$ & $0.0619021$  & $0.119486$   &  $49.759$  \\
		$\beta$ Pic b ($i=2$) & $46.6215$  & $0.119486$  & $-0.0631892$ &$125.854$ \\
		\hline
	\end{tabular}
    \label{eigen1}
\end{table}

I find that the apsidal eigenfrequency for $\beta$ Pic c and $\beta$ Pic b are $15.7^{\prime \prime} \rm yr^{-1}$ and $46.6^{\prime \prime} \rm yr^{-1}$, respectively. The outermost planet, $\beta$ Pic b, has the largest apsidal eigenfrequency, while the innermost planet, $\beta$ Pic c, has the lowest. The resulting components of the eigenvectors, $\bar{e}_{ji}$, from the apsidal eigenfrequencies are initially unscaled. I scale the components of the eigenvectors  so that
\begin{equation}
S_i \bar{e}_{ji} = e_{ji},
\end{equation}
where $S_i$ denotes the scaling constant \citep[see Chapter $7.3$ of][]{MurrayBook2000}. I define the initial vertical and horizontal components of the eccentricity vectors with
\begin{equation}
h_j = e_j \sin(\bar{\omega}_j)
\end{equation}
and
\begin{equation}
k_j = e_j \cos(\bar{\omega}_j),
\end{equation}
where $h_j$ and $k_j$ can be determined by the initial conditions shown in Table~\ref{table::planets} (using the parameters from \cite{Brandt2021}). The vertical and horizontal components at $t \neq 0$ are defined by
\begin{equation}
h_j(t) = \sum_{i=1}^N e_{ji} \sin(g_i t + \beta_i)
\label{eqn_j}
\end{equation}
and
\begin{equation}
k_j(k) = \sum_{i=1}^N e_{ji} \cos(g_i t + \beta_i),
\label{eqn_k}
\end{equation}
where $\beta_i$ represents the phase angle. I therefore have two sets of two simultaneous linear equations, from equations~(\ref{eqn_j}) and~(\ref{eqn_k}), with four unknowns, $S_i \sin \beta_i$ and $S_i \cos \beta_i$, with $i = 1,2$. By solving these linear equations, I can find a value for both the scaling constants $S_i$ and the phases $\beta_i$. The scaled eigenvector components along with the phase angles are shown in columns $3-4$ of Table~\ref{eigen1}.

\begin{figure}
\includegraphics[width=\columnwidth]{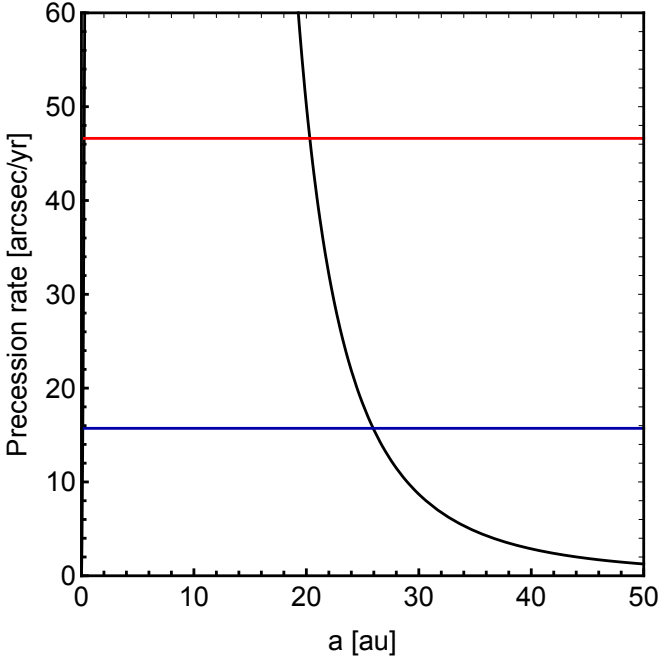}
\caption{Test particle free precession rate (solid black line) given by $g_0$ in equation~(\ref{Aprec}) as a function of semi--major axis. The apsidal eigenfrequencies of each of the planets are given by the horizontal lines. The eigenfrequencies are those of $\beta$ Pic b (red), and $\beta$ Pic c (blue). Apsidal secular resonances are located where the free precession rate of the test particle equals the apsidal eigenfrequency of a planet.}
\label{fig::prec}
\end{figure}

\subsection{Planetesimal Free Precession Rate}
\label{precrate}

I now calculate the free precession rate of a test particle in the potential of the $\beta$ Pic planetary system. The free precession rate is given by
\begin{equation}
g_0 = \frac{n}{4}\sum_{j=1}^N\frac{m_j}{m_*}\alpha_j \bar{\alpha}_j b_{3/2}^{(1)}(\alpha_j),
\label{Aprec}
\end{equation}
where $n$ is the orbital frequency of the test particle \citep[e.g.][]{MurrayBook2000}. The variables $\alpha_j$ and $\bar{\alpha}_j$ are defined as
\begin{equation}
  \alpha_{j}=\begin{cases}
    a_j/a, & \text{if $a_j < a$},\\
    a/a_j, & \text{if $a_j > a$},
  \end{cases}
\end{equation}
and 
\begin{equation}
  \bar{\alpha}_{j}=\begin{cases}
    1, & \text{if $a_j < a$},\\
    a/a_j, & \text{if $a_j > a$},
  \end{cases}
\end{equation}
where $a$ is the semimajor axis of the test particle.

The solid line in Fig.~\ref{fig::prec} shows the free precession rate of a test particle as a function of semimajor axis. The horizontal lines show the apsidal eigenfrequencies of the two planets. The intersection of the test particle's free precession rate with an apsidal eigenfrequency marks the location of an apsidal secular resonance \citep{Minton2011,Haghighipour2016,Smallwood2018a,Smallwood2018b,Smallwood2021}. There are two such secular resonances that arise exterior to the orbit of $\beta$ Pic b at $\sim 20\, \rm au$ and $\sim 25\, \rm au$. I denote the location of the innermost intersection of the test particle's free precession rate with the eigenfrequency of $\sim 46^{\prime \prime} \rm yr^{-1}$ as the $\nu_{\rm b}$ secular resonance, while the outermost intersection as the the $\nu_{\rm c}$ secular resonance.

\subsection{Eccentricity Excitation}
To determine the strength of each of the two secular resonances in the $\beta$ Pic planetary system, I compute the forced eccentricity of a test particle as a function of semimajor axis. If the forced eccentricity is large, debris may be ejected from the system or collide with a larger object. I begin with the secular resonant disturbing function, $\mathcal{R}^{\rm sec}$,  from \cite{MurrayBook2000} that describes the secular theory for $N$ planets including a test particle with mean motion, $n$, eccentricity, $e$, inclination, $I$, and longitude of the perihelion, $\bar{\omega}$, given by
\begin{align}
\mathcal{R}^{\rm sec} =  na^2\bigg[ \frac{1}{2}g_0e^2 + \frac{1}{2}BI^2 & + \sum_{j=1}^N A_j e e_j \cos(\bar{\omega} - \bar{\omega}_j)
\notag \\
& + \sum_{j=1}^N B_j I I_j \cos(\Omega - \Omega_j)\bigg].
\label{R}
\end{align}
Since I consider a coplanar system, equation~(\ref{R}) can be simplified to include only the terms involving the eccentricity
\begin{equation}
\mathcal{R}_{\rm ecc}^{\rm sec} = na^2 \bigg[\frac{1}{2}g_0 e^2  + \sum_{j=1}^N A_j e e_j \cos(\bar{\omega} - \bar{\omega}_j)\bigg],
\end{equation}
where $g_0$ is the test particle free precession rate given in equation~(\ref{Aprec}) and
\begin{equation}
A_j = -n \frac{1}{4}\frac{m_j}{m_{*}}\alpha_j \bar{\alpha}_j b_{3/2}^{(1)}(\alpha_j).
\end{equation}
The forced eccentricity is given by
\begin{equation}
e_{\rm forced} = \sqrt[]{h_0^2(t) + k_0^2(t)},
\label{forced}
\end{equation}
where
\begin{equation}
h_0(t) = -\sum_{i=1}^N\frac{\nu_i}{g_0-g_i}\sin(g_i t + \beta_i)
\end{equation}
and
\begin{equation}
k_0(t) = -\sum_{i=1}^N\frac{\nu_i}{g_0-g_i}\cos(g_i t + \beta_i).
1\end{equation}
The constants $\beta_i$ are determined from the initial boundary conditions (see Table~\ref{eigen1}), and  $\nu_i$ is given by
\begin{equation}
\nu_i = \sum_{j=1}^N A_j e_{ji},
\label{nui}
\end{equation}
where $e_{ji}$ are the scaled eigenvector components corresponding to the eigenfrequencies calculated using equations~(\ref{A1}) and~(\ref{A2}). 

\begin{figure}
\includegraphics[width=\columnwidth]{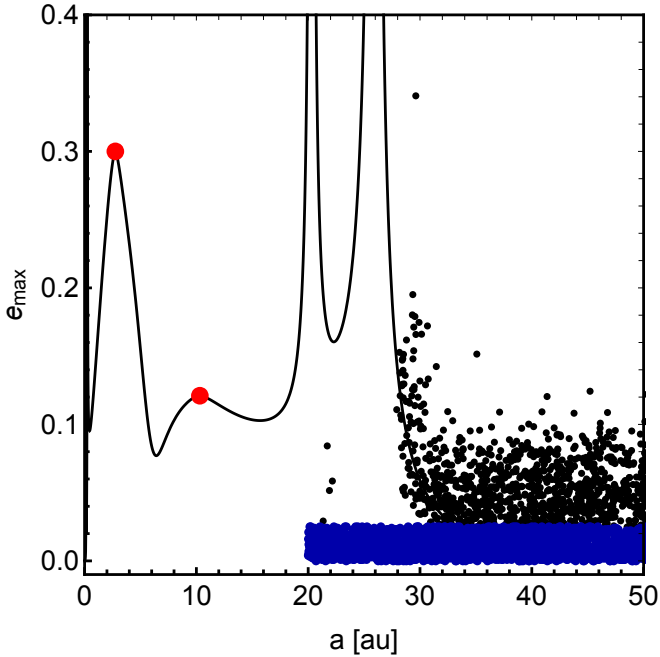}
\caption{The maximum forced eccentricity as a function of semi-major axis of a test particle, showing the eccentricity excitation region for the two secular resonances centered at $\sim 20\, \rm au$ and $\sim 25\, \rm au$. The semi-major axis and eccentricity of the $\beta$ Pic b and $\beta$ Pic c are given by the red dots. The blue dots are the initial distribution of the $N$--body simulation and the black dots are the particle distribution after $20\, \rm Myr$. The secular resonances are perturbing particles within $\sim 25\, \rm au$, which are responsible for creating the inner edge of the debris disc in $\beta$ Pic.}
\label{fig::sec_res}
\end{figure}

Figure~\ref{fig::sec_res} shows the forced eccentricity of a test particle given by equation~(\ref{forced}) as a function of semi-major axis at time $t = 0$, which corresponds to present day conditions. Each of the planets' eccentricities are denoted by the red dots. The wider the region with high forced eccentricity, the more planetesimals can potentially undergo secular resonant perturbations. In this figure, I also include the initial particle distribution and the distribution at $t = 20\, \rm Myr$ of my  $N$--body simulation. The particles are cleared out near the location of the analytically derived secular resonances. Therefore, the secular resonances are responsible for shaping the inner edge of the $\beta$ Pic debris disc.

\section{Conclusions}
\label{sec::conclusions}
In this work, I modelled a protoplanetary gas disc and a debris disc under the influence of the known outer planet, $\beta$ Pic b, and the newly confirmed inner planet, $\beta$ Pic c.  Unlike previous simulations of this system, I use the updated planetary system parameters from \cite{Brandt2021}. Observations revealed a strong  warp of the debris disc, where the 'inner disc' is inclined by $4-5\degree$ with respect to the outer portions of the 'outer disc.' My hydrodynamical simulations reveal that inclined planets do not perturb the gas disc in such a way as to produce a  long lasting warp,  even when modeling the observed disc size. The warp will propagate across the entire disc with a timescale that is much less than the  gas disc lifetime. Therefore, the observed warp must be generated after the gas disc disperses.  With $N$--body simulations, I found that the inner debris disc edge is truncated when the two planets are included. Since both planets have a nonzero eccentricity, I found that two secular resonances are present exterior to the orbit of $\beta$ Pic b. These secular resonances cause the clearing of material from $20-25\, \rm au$, which is responsible for truncating the inner edge of the $\beta$ Pic debris disc.

\section*{Acknowledgements}
 JLS thanks the anonymous referee for helpful suggestions that positively impacted the work. JLS acknowledges funding from the ASIAA Distinguished Postdoctoral Fellowship. JLS thanks Rebecca G. Martin, Lorin Matthews, and Ruobing Dong for insightful discussions that improved the manuscript's quality.

\section*{Data Availability}

The data supporting the plots within this article are available on reasonable request to the corresponding author. A public version of the {\sc phantom} and {\sc mercury} codes are available at \url{https://github.com/danieljprice/phantom} and \url{https://github.com/4xxi/mercury}, respectively.



\bibliographystyle{mnras}
\bibliography{ref.bib} 




\appendix

\section{Planet-disc interaction}
\label{appendix_a}

\begin{figure*}
\includegraphics[width=2\columnwidth]{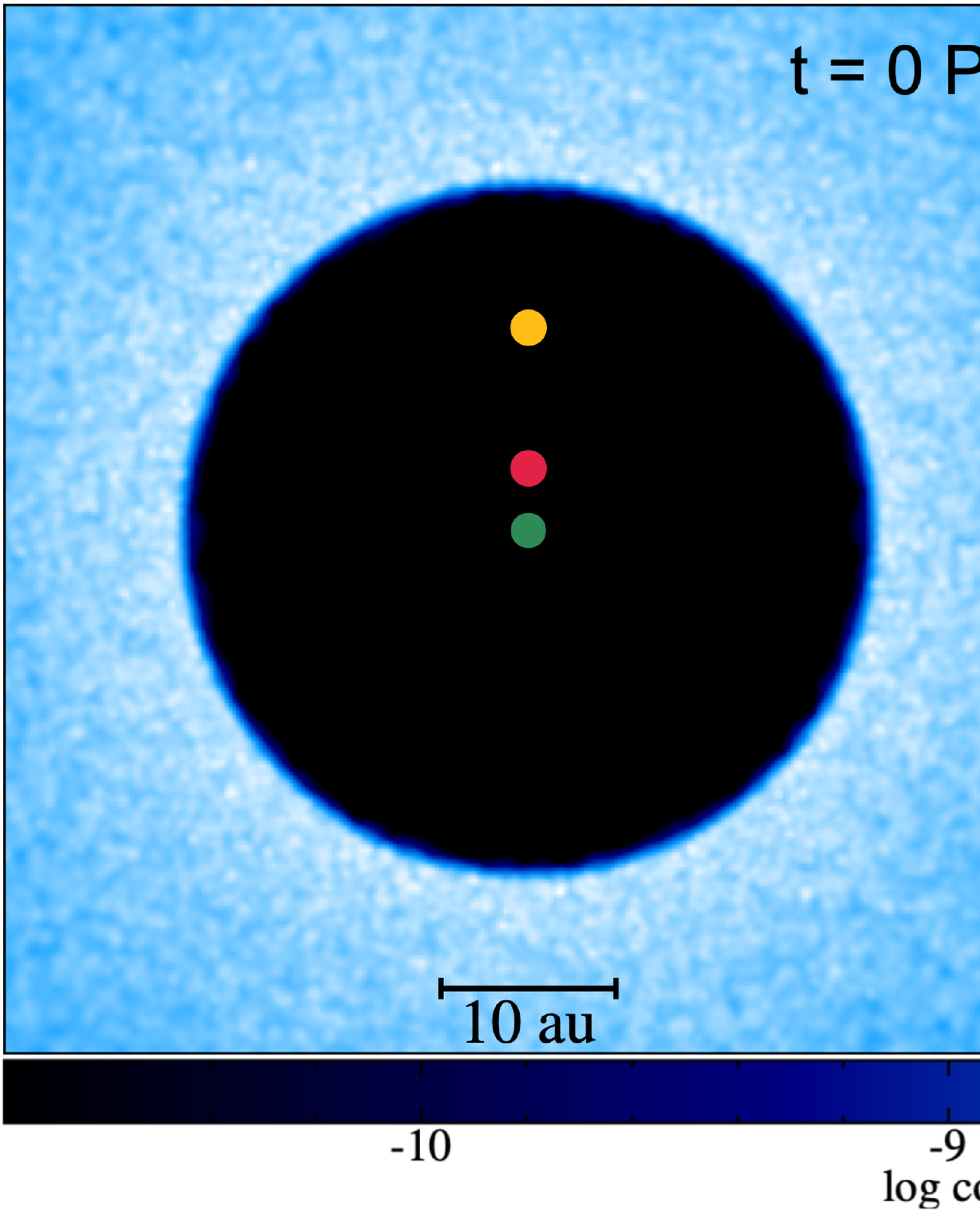}
\caption{ The interaction between the disc and the $\beta$ Pic planets for $\gamma_{\rm p} = 20^\circ$. The left panel shows the disc structure at $t = 0\, \rm P_b$. The green dot denotes the central star, while the yellow and red dots denote planets b and c, respectively. The right panel shows the disc structure at $t = 1000\, \rm P_b$. The inner edge of the disc interacts with the planets.}
\label{fig::splash}
\end{figure*}

 In the hydrodynamical simulations, the planets are initially interior to the initial disc edge. However, the inner edge of the disc will viscously spread inwards, eventually interacting with the planets. The left panel in Fig.~\ref{fig::splash} shows the initial setup where $\beta$ Pic b (yellow dot) and $\beta$ Pic c (red dot) are initially interior to the inner disc edge. The right panel in Fig.~\ref{fig::splash} shows the disc structure at a $t = 1000\, \rm P_b$. At this time, the inner edge of the disc interacts with the planets.




\section{Planetary evolution}
\label{appendix_c}


\begin{figure*}
\includegraphics[width=2\columnwidth]{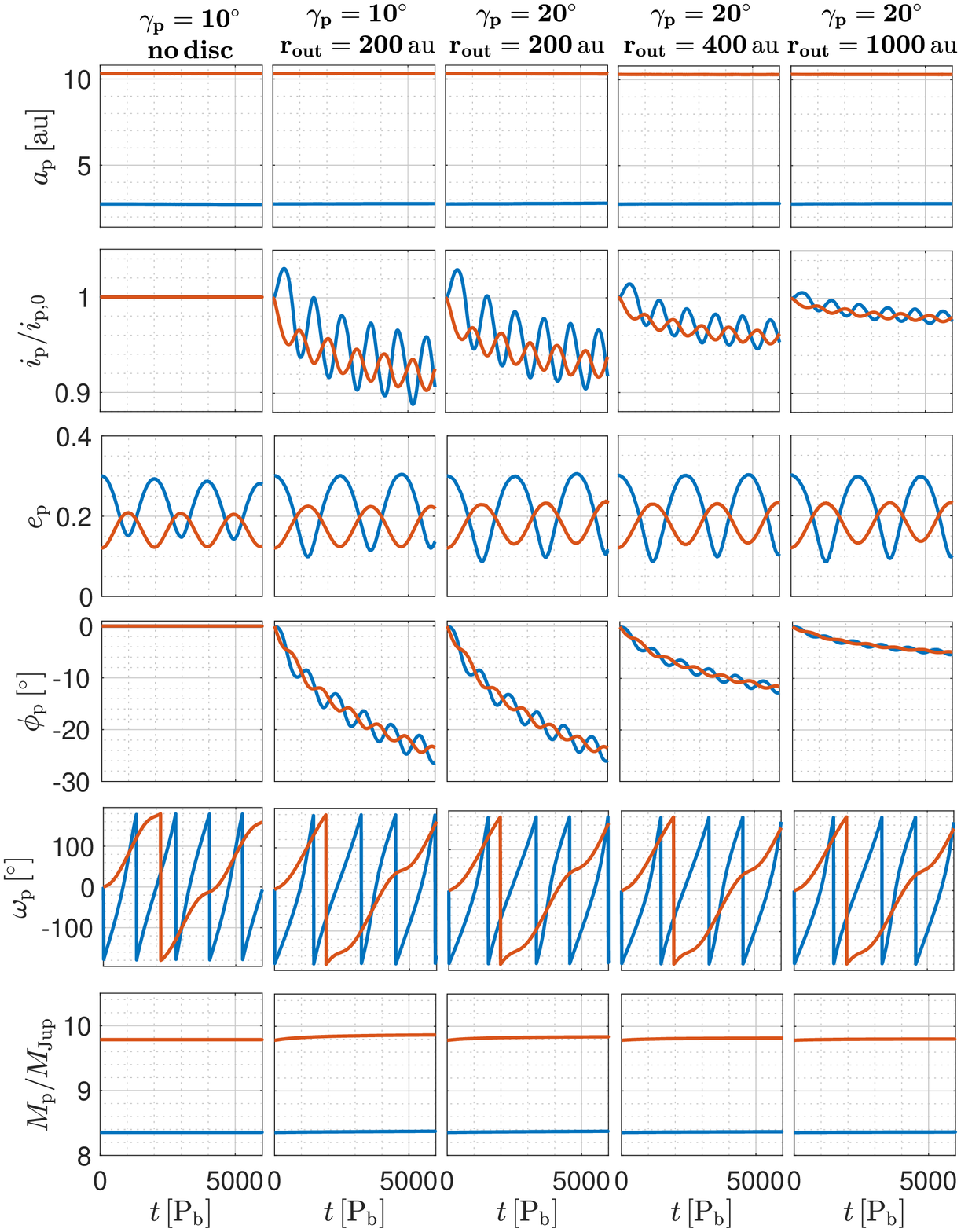}
\caption{Evolution of the  orbits of the two $\beta$ Pic planets from the hydrodynamic simulations. The initial simulation parameters are given at the top of each column.  I show the evolution of the semi-major axis ($a_{\rm p}$, row 1),  tilt ($i_{\rm p}/i_{\rm p,0}$, row 2), eccentricity ($e_{\rm p}$, row 3), the longitude of the ascending node ($\phi_{\rm p}$, row 4), the argument of the pericentre ($\omega_{\rm p}$, row 5), and planetary mass ($M_{\rm p}$, row 6) as a function of time in orbital periods of the outer-most planet ($\beta$ Pic b).  The blue and  red lines represent the planetary parameters for $\beta$ Pic c and $\beta$ Pic b, respectively. }
\label{fig::planet_params}
\end{figure*}


 I examine the evolution of the two eccentric planets, $\beta$ Pic b and $\beta$ Pic c, in all hydrodynamical simulations. Fig.~\ref{fig::planet_params} shows the evolution of the semi-major axis ($a_{\rm p}$, row 1), tilt ($i_{\rm p}$, row 2), eccentricity ($e_{\rm p}$, row 3), longitude of the ascending node ($\phi_{\rm p}$, row 4), the argument of the pericentre ($\omega_{\rm p}$, row 5), and planetary mass ($M_{\rm p}$, row 6) as a function of time. Each column denotes a different simulation with the initial parameters given at the top of each column. The left-most column shows a simulation with no gas disc present with initial planetary tilts $i_{\rm p} = 10^\circ$. The only significant difference in the planetary parameters is that there are no tilt oscillations since there is no gas disc perturbing the planetary orbits. In each case,  both planets remain at nearly their initial separation from the central star. The planets undergo tilt oscillations driven by the gas disc, which are out of phase with one another. As the angular momentum of the disc increases (larger disc sizes), the tilt of the planetary orbits decreases at a slower rate as they align to the total angular momentum of the system. The eccentricities of the inner and outer planets also oscillate in time, with the amplitude of the outer planet's eccentricity larger than the inner planet. Since the model is for a low-mass disc and the change in planetary tilt is small, the planets precess extremely slowly. However, as shown in the fifth panel, the eccentric planets undergo apsidal precession. The inner planet apsidally precesses faster than the outer planet. Finally,  the masses of planets b and c increase slowly over time due to the accretion of material.



\bsp	
\label{lastpage}
\end{document}